\documentclass[]{aastex631}
\usepackage{verbatim}
\usepackage{booktabs}

\begin{document}
\newcommand{\tSrc}{1ES 1959+650}

%\title{A Multi-wavelength studies of TeV flares from 1ES 1959+650 triggered by LHAASO-WCDA}
\title{LHAASO-WCDA observed a $\sim$ 5 days TeV-delayed flaring event in blazar \tSrc}
\correspondingauthor{G.M. Xiang, J.N. Zhou, L. Chen, M. Zha}
\email{gmxiang@ihep.ac.cn, zjn@shao.ac.cn, chenliang@shao.ac.cn, zham@ihep.ac.cn}

%\correspondingauthor{G.M.XIANG, J.N.ZHOU, LIANG.Chen, M.ZHA}



 
\author{Zhen Cao}
\affiliation{State Key Laboratory of Particle Astrophysics \& Experimental Physics Division \& Computing Center, Institute of High Energy Physics, Chinese Academy of Sciences, 100049 Beijing, China}
\affiliation{University of Chinese Academy of Sciences, 100049 Beijing, China}
\affiliation{Tianfu Cosmic Ray Research Center, 610000 Chengdu, Sichuan,  China}
 
\author{F. Aharonian}
\affiliation{Tianfu Cosmic Ray Research Center, 610000 Chengdu, Sichuan,  China}
\affiliation{University of Science and Technology of China, 230026 Hefei, Anhui, China}
\affiliation{Yerevan State University, 1 Alek Manukyan Street, Yerevan 0025, Armenia }
\affiliation{Max-Planck-Institut for Nuclear Physics, P.O. Box 103980, 69029  Heidelberg, Germany}
 
\author{Y.X. Bai}
\affiliation{State Key Laboratory of Particle Astrophysics \& Experimental Physics Division \& Computing Center, Institute of High Energy Physics, Chinese Academy of Sciences, 100049 Beijing, China}
\affiliation{Tianfu Cosmic Ray Research Center, 610000 Chengdu, Sichuan,  China}
 
\author{Y.W. Bao}
\affiliation{Tsung-Dao Lee Institute \& School of Physics and Astronomy, Shanghai Jiao Tong University, 200240 Shanghai, China}
 
\author{D. Bastieri}
\affiliation{Center for Astrophysics, Guangzhou University, 510006 Guangzhou, Guangdong, China}
 
\author{X.J. Bi}
\affiliation{State Key Laboratory of Particle Astrophysics \& Experimental Physics Division \& Computing Center, Institute of High Energy Physics, Chinese Academy of Sciences, 100049 Beijing, China}
\affiliation{University of Chinese Academy of Sciences, 100049 Beijing, China}
\affiliation{Tianfu Cosmic Ray Research Center, 610000 Chengdu, Sichuan,  China}
 
\author{Y.J. Bi}
\affiliation{State Key Laboratory of Particle Astrophysics \& Experimental Physics Division \& Computing Center, Institute of High Energy Physics, Chinese Academy of Sciences, 100049 Beijing, China}
\affiliation{Tianfu Cosmic Ray Research Center, 610000 Chengdu, Sichuan,  China}
 
\author{W. Bian}
\affiliation{Tsung-Dao Lee Institute \& School of Physics and Astronomy, Shanghai Jiao Tong University, 200240 Shanghai, China}
 
\author{J. Blunier}
\affiliation{APC, Universit'e Paris Cit'e, CNRS/IN2P3, CEA/IRFU, Observatoire de Paris, 119 75205 Paris, France}
 
\author{A.V. Bukevich}
\affiliation{Institute for Nuclear Research of Russian Academy of Sciences, 117312 Moscow, Russia}
 
% ...
\collaboration{334}{The LHAASO Collaboration}

\begin{abstract}

%We report a day-scale hard lag between GeV and TeV $\gamma$-ray emission from the HBL 1ES~1959+650 in early 2024. \textcolor{red}{Since} the LHAASO-WCDA real-time monitoring system began operation in late 2023, multiple TeV flares from this source \textcolor{red}{have been triggered, including the 1st trigger flare on 2024 February 9}.
%A Bayesian-block analysis of the WCDA light curve identifies three TeV flares in 2024. For the second triggered \textbolor{red}{flare}, a discrete cross-correlation analysis yields a TeV delay of \textcolor{red}{$\Delta t = 6.0_{-2.9}^{+3.0}$} days relative to the GeV flare, with Monte Carlo red-noise simulations indicating a significance of \textcolor{red}{3.8 $\sigma$}. Time-resolved spectroscopy shows that this event has the softest TeV spectrum among \textcolor{red}{these flares} (intrinsic spectral index $\Gamma=3.16\pm0.18$), while the 1st trigger flare is harder ($\Gamma=2.48\pm0.21$). The observed day-scale hard lag favors an acceleration-dominated evolution (e.g., stochastic/Fermi-II acceleration) rather than purely radiative cooling, and constrains particle energization and the location/kinematics of the $\gamma$-ray emitting region in the jet.

We report a day-scale hard lag between GeV and TeV $\gamma$-ray emission from the HBL 1ES~1959+650 in early 2024. Since the LHAASO-WCDA real-time monitoring system began operation in late 2023, multiple TeV flares from this source have been triggered, including the 1st trigger flare on 2024 February 9.
A Bayesian-block analysis of the WCDA light curve identifies three TeV flares in 2024. For the second triggered flare, a discrete cross-correlation analysis reveals a $>3\,\sigma$ correlation (relative to uncorrelated red-noise simulations) at a time delay of $\Delta t = 5.0_{-2.1}^{+2.1}$ days, with the TeV emission lagging the GeV.
Time-resolved spectroscopy shows that this flare has the softest TeV spectrum among these flares (intrinsic spectral index $\Gamma=3.16\pm0.18$), while the 1st trigger flare is harder ($\Gamma=2.48\pm0.21$). The observed five-day hard lag is difficult to reconcile with a purely cooling-driven temporal ordering and is consistent with scenarios in which particle energization and/or transport may contribute to the evolution. However, the current data do not uniquely identify the underlying mechanism.

\end{abstract}


%\maketitle


\section{Introduction} \label{sec:intro}

Blazars, a subclass of active galactic nuclei (AGN), are among the most extreme and energetic objects in the universe. Their relativistic jets, oriented close to the observer's line of sight, produce non-thermal emission across the entire electromagnetic spectrum from radio to very high-energy (VHE) gamma rays \citep{1995PASP..107..803U}. As such, blazars serve as natural laboratories for studying particle acceleration, relativistic jet physics, and the extragalactic background light (EBL) \citep{2019Galax...7...20B}. A key observational handle is their pronounced variability over a wide range of timescales. In particular, correlated multi-band variability and inter-band time lags can directly probe the competition between acceleration, cooling, and transport processes, as well as the geometry and evolution of the emitting region. However, robust lag detections at $\gamma$-ray energies are still relatively rare, especially when the lag extends across widely separated bands (e.g., GeV to TeV), making such events potentially discriminating diagnostics of the underlying variability physics.

The gamma-ray emission in blazars is widely attributed to non-thermal processes involving relativistic electrons and, in some cases, hadrons within the jet \citep{2011JApA...32..139U,2019Galax...7...20B}. Leptonic models explain the high-energy emission through inverse Compton scattering, where relativistic electrons upscatter low-energy photons (e.g., synchrotron or external radiation) to gamma-ray energies. Alternatively, hadronic models propose that gamma rays are produced via proton synchrotron radiation or the decay of neutral pions resulting from proton--photon interactions. These mechanisms are often accompanied by the production of high-energy neutrinos, making blazars promising candidates for multimessenger astrophysics.

1ES 1959+650 is a high-frequency-peaked BL Lacertae (HBL) object, located at a redshift of $z=0.048$. It has garnered particular attention not only for its intense gamma-ray activity but also for its potential role as a neutrino emitter. A comprehensive 2002 target-of-opportunity multiwavelength campaign, triggered by strong TeV activity, revealed multiple pronounced flares and enabled sensitive tests of the X-ray--TeV flux correlation: while the X-ray and TeV fluxes appeared broadly correlated overall, an ``orphan'' TeV flare was identified without a contemporaneous X-ray counterpart, with no compelling evidence for optical correlations and with radio emission remaining approximately steady within uncertainties \citep{2004ApJ...601..151K}. The same campaign also highlighted that simple one-zone SSC descriptions are challenged by such episodes and by the comparatively weak radio/optical variability relative to X-ray/TeV variability, motivating more complex (e.g., multi-zone or structured) emission pictures in at least some states \citep{2004ApJ...601..151K}. Building on this historical context, \citep{2013PhRvD..87j3015S} discussed the orphan-flare episode as a potential site of enhanced high-energy neutrino production. Later, the AMANDA collaboration claimed the detection of two neutrinos coincident with TeV flares \citep{2005APh....23..537H}; however, their statistical significance cannot be reliably estimated since the events were not detected in a blind analysis. From May to July 2016, \tSrc~was observed experiencing strong gamma-ray activity. The IceCube collaboration performed a search for neutrinos coincident with the flares, but without a significant excess being observed \citep{2017ICRC...35..969K}.

From January 2024, 1ES 1959+650 entered a gamma-ray reactivated state, shortly after the commissioning of the real-time monitoring system based on the Water Cherenkov Detector Array of the Large High Altitude Air Shower Observatory(LHAASO-WCDA) \citep{2024RAA....24l5020X}. The first TeV flaring event was triggered on February 9th, 2024 \citep{2024ATel16437....1X}. 
We then conduct a comprehensive multi-wavelength study by combining LHAASO-WCDA monitoring with contemporaneous data from Fermi-LAT and X-ray instruments (MAXI and Swift). By focusing on the time-domain evolution, specifically investigating inter-band temporal correlations, and performing broadband spectral energy distribution (SED) modeling, we seek to provide robust constraints on the particle acceleration and emission processes underlying these flares.
The paper is organized as follows. Section 2 details the multi-wavelength data analysis process. In Section 3, we investigate the temporal and spectral characteristics of the flaring episodes. Finally, Section 4 discusses the plausible astrophysical interpretations and implications of our findings.

\section{Data analysis} \label{sec:data}

\subsection{LHAASO-WCDA Observation}
The LHAASO (Large High Altitude Air Shower Observatory) is a hybrid, large-area array with a wide field of view, designed for investigating cosmic rays (CRs) and gamma rays across a broad energy spectrum, from sub-TeV to PeV. It is located on Haizi Mountain, about 4410 meters above sea level in Daocheng, Sichuan province, China. It incorporates three primary detector arrays: the 78,000 m$^2$ Water Cherenkov Detector Array (WCDA), the 1.3 km$^2$ Kilometer Square Array (KM2A), and the Wide Field-of-view Cherenkov Telescope Array (WFCTA). LHAASO began partial operations in April 2019 and entered full scientific operation in July 2021 \citep{cao2022largehighaltitudeair}. 
The WCDA is optimized for detecting gamma rays in the energy range from 100 GeV to 25 TeV\citep{Crab_WCDA,cao2022largehighaltitudeair}, while the KM2A array is sensitive to energies from roughly 25 TeV to several PeV \citep{CPC2021,cao2022largehighaltitudeair}.

The WCDA reconstructed events are sorted into 7 $N_{\text{hit}}$ bins, defined by the effective hit multiplicity of PMTs, [30–60), [60–100), [100–200), [200–300), [300–500), [500–800), and [800–2000], spanning median energies from $\sim0.6$ to $\sim$~25 TeV, serves as a rough proxy of the energy of the primary particle.
For each analysis bin, a gamma/proton separation parameter is applied to reduce the number of hadronic background events. A full description of the WCDA detector and its reconstruction procedures can be found in \citep{Crab_WCDA}. Event and background maps were generated in celestial coordinates (Epoch J2000.0) with a pixel size of 0.1$^{\circ}$ × 0.1$^{\circ}$. 
The cosmic-ray background in each pixel is estimated using the direct integration method \citep{Fleysher_2004}, with an integration time of 10 hours. To ensure the purity of the background estimation, regions within the Galactic plane ($|b| < 10^{\circ}$) and known gamma-ray sources (with a spatial separation less than 5$^{\circ}$) were masked from the calculation.
Since EBL absorption becomes progressively more severe at higher energies, for this study, we use only WCDA data and do not include data from KM2A. The data collection period spans from 8 March 2021 to 31 July 2024, totaling 1241 days of effective live time. To ensure the quality of reconstruction, only events with a zenith angle of $\theta < 50^{\circ}$ are selected in this analysis. The analysis employs the standard LHAASO-WCDA three-dimensional maximum-likelihood software package. Details regarding the light curve and spectral energy distribution (SED) are provided in the Appendix.

\subsection{Multi-wavelength Observations}
We analyzed multi-wavelength data from gamma-ray, X-ray, and optical instruments to obtain a comprehensive temporal and spectral characterization of \tSrc.

{\bf Fermi-LAT} — We used 0.1–300 GeV PASS 8 SOURCE class events from a $20^{\circ}\times 20^{\circ}$ ROI centered on \tSrc. The data span matches that of LHAASO-WCDA. The analysis employed Fermitools v2.2.0 with IRF P8R3\_SOURCE\_V3. The source model includes all 4FGL-DR4 catalog sources \citep{2020ApJS..247...33A} and the standard Galactic (gll\_iem\_v07.fits) and isotropic (iso\_P8R3\_SOURCE\_V3\_v1.txt) diffuse components. Background parameters were fixed, while variable sources and those within $5^\circ$ of \tSrc~had free normalizations. We model the source spectrum with a log-parabola following the catalog. Light curves were computed in 1- and 3-day bins, with 90\% upper limits quoted for points having TS$<$5. For spectral energy distribution (SEDs), the 0.1–300 GeV range was divided into six logarithmic energy bins.

{\bf X-ray Observations} — We used X-ray monitoring data from MAXI and Swift. MAXI provides 2–20 keV all-sky coverage \citep{2009PASJ...61..999M}; we used daily-averaged light curves from MJD 60157–60523. Swift-BAT supplies 15–150 keV hard X-ray monitoring, from which we obtained the daily light curve from the public archive. Swift-XRT provides 0.3–10 keV pointed observations; a total of 48 XRT observations were available within the same period. The XRT data were processed using the on-demand analysis tools \citep{2009MNRAS.397.1177E}, and the resulting light curve was binned into 1-day intervals.
We fitted the spectra with an absorbed power-law model ({\tt TBabs*zTBabs*powerlaw}) and derived the intrinsic flux using the {\tt cflux} convolution model. To construct the SED, we divided the 0.1–10.0 keV XRT data into three equal logarithmic energy bins.

{\bf Optical} — Optical monitoring data were obtained from the ASAS-SN public archive \citep{2014AAS...22323603S}, which provides long-term V/g-band light curves with full-sky coverage down to $\sim17$ mag.

%\begin{figure}
%\plotone{WCDA_LC.pdf}
%\caption{LHAASO-WCDA Light curve of 1ES 1959+650, binned in 7-day intervals. Active and quiescent periods, determined using Bayesian methods, are indicated by the dashed red line. }
%\label{fig:WCDA_LC}
%\end{figure}

%% The "ht!" tells LaTeX to put the figure "here" first, at the "top" next
%% and to override the normal way of calculating a float position
\begin{figure}
\plotone{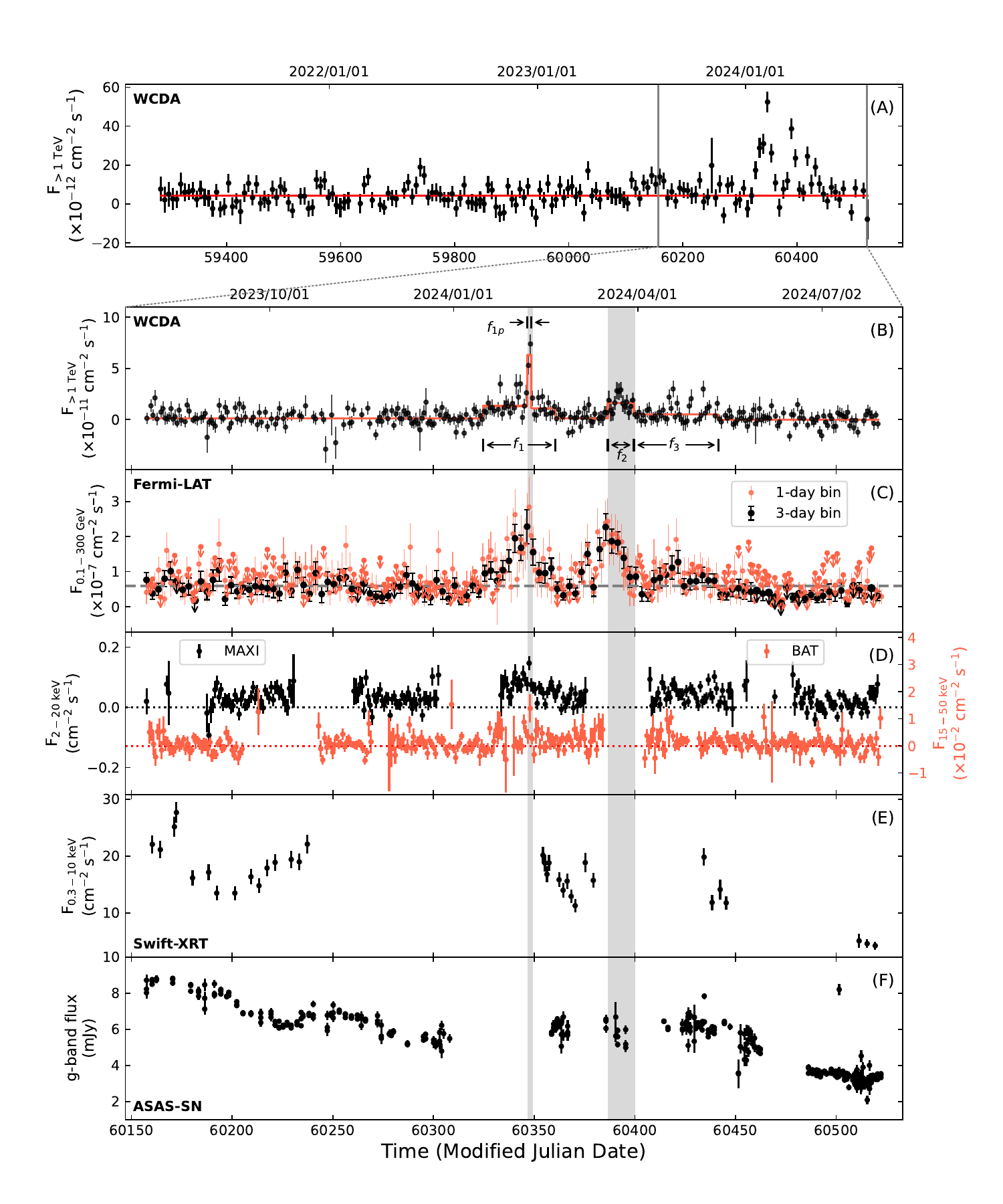}
\caption{Multi-wavelength light curves of \tSrc. The gray vertical lines in Panel (A) indicate the time interval expanded in Panels (B)–(F). The two gray shaded regions mark the periods that triggered the LHAASO-WCDA monitoring system. {\bf (A)} Long-term LHAASO-WCDA light curve in weekly bins. The red line represents the mean flux level during the quiescent state. Three distinct flaring events are identified: flare $f_1$ (MJD 60324.7--60360.6), flare $f_2$ (MJD 60386.6--60399.5), and flare $f_3$ (MJD 60399.5--60441.4).  The $f_{1p}$ (MJD 60346.677--60348.671) is the first triggered event reported by the LHAASO-WCDA monitoring system. {\bf (B)} LHAASO-WCDA light curve binned in 1-day intervals. The red solid line represents the Bayesian block representation used to define active and quiescent states. {\bf (C)} Fermi-LAT light curves in 1-day (red) and 3-day (black) bins. The horizontal dashed line denotes the mean flux level. {\bf (D)} Daily-binned X-ray monitoring data from MAXI (2–20 keV, left axis, black) and Swift-BAT (15–50 keV, right axis, red). {\bf (E)} Swift-XRT (0.3–10 keV) light curve, rebinned to 1-day intervals. {\bf (F)} Optical g-band light curve from ASAS-SN.}
\label{fig:MLC}
\end{figure}

\begin{figure}
  \centering
  \includegraphics[width=0.45\textwidth]{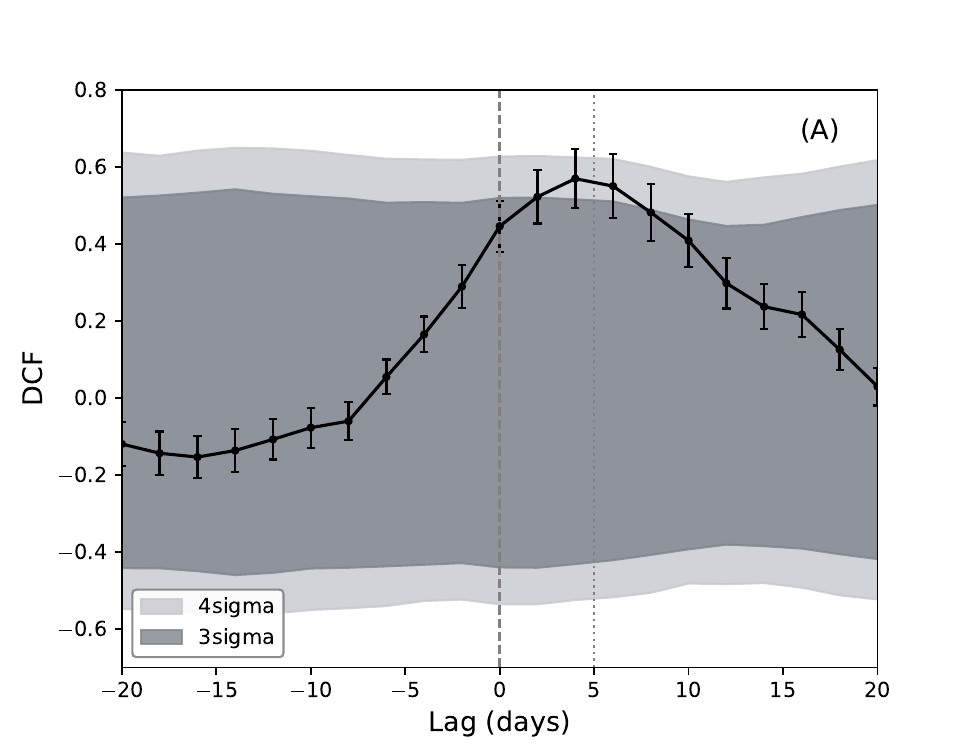}
  \includegraphics[width=0.45\textwidth]{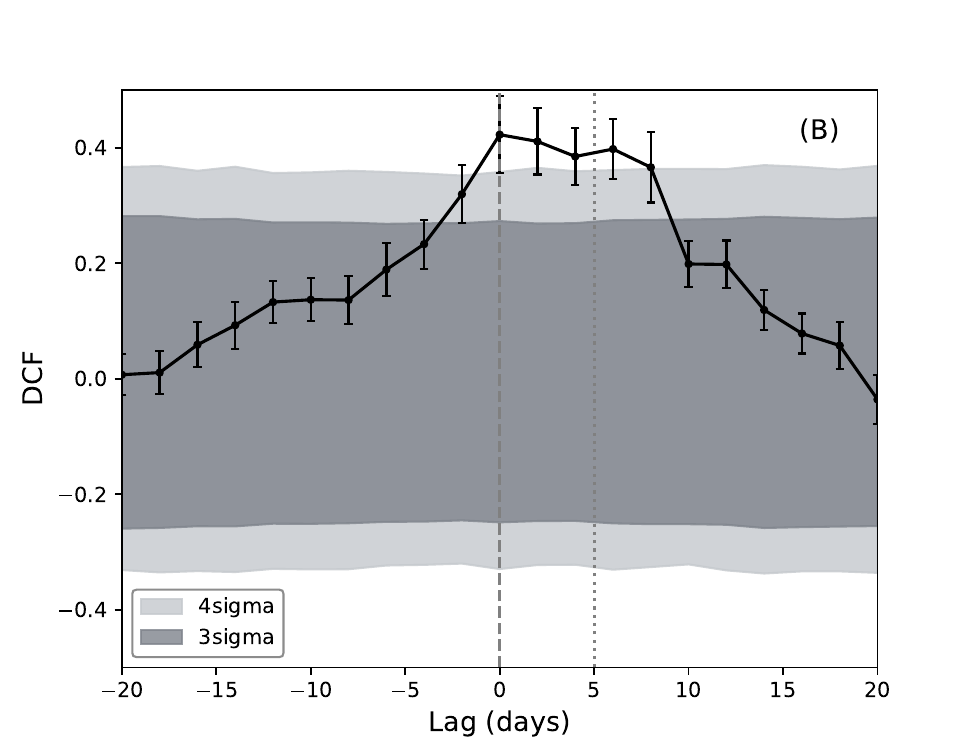}
  \caption{Discrete cross-correlation functions between the TeV and GeV light curves. {\bf (A)}Light curve from MJD 60300 to 60500, excluding Flare 1 and Flare 3. {\bf (B)} Full light curve, corresponding to panel (B) of Figure~\ref{fig:MLC}. Dashed and dotted lines mark lags of 0 and 5.0 days, respectively.}
  \label{fig:DCF}
\end{figure}

%\begin{figure}
%\plotone{GeV_TeV_SED.pdf}
%\caption{Spectral energy distributions (SEDs) of \tSrc~during different periods from GeV to TeV energies. Periods are defined according the TeV flux shown in Fig.~\ref{fig:MLC}.}
%\label{fig:SED}
%\end{figure}

\subsection{Correlation analysis}  \label{simu}
To quantify the correlation and potential time lag between the light curves of the VHE and GeV bands, we applied the Discrete Cross-Correlation Function (DCF) method \citep{1988ApJ...333..646E}. 
The DCF is specifically designed for astrophysical time series that are characterized by uneven sampling and heteroscedastic noise. 
The primary objective is to evaluate whether the flux changes in different energy bands are synchronous or if there exists a characteristic time lag ($\Delta t$), which provides critical constraints on the underlying radiation mechanisms and the spatial distribution of emitting regions.

%The uncertainty of the time lag was estimated using the Flux Randomization and Random Subset Selection (FR/RSS) Monte Carlo method \citep{Peterson_1998}. Specifically, we modulated the flux values by adding Gaussian noise to each data point, with the noise width equal to the observed measurement uncertainty. For each iteration, a subset containing approximately $63.3\%$ of the original data points was randomly selected to account for sampling effects. A total of $10,000$ Monte Carlo realizations were generated for both energy bands. For each pair of simulated light curves, the DCF was calculated, and the centroid lag ($\tau_{\text{cen}}$) was determined. The centroid was defined as the weighted average of the time lags where the DCF coefficients exceeded $80\%$ of the peak value. This process yielded a distribution of centroid lags; the median of this distribution was adopted as the best estimate for the time lag, while the 16th and 84th percentiles were used to define the 1 $\sigma$ statistical uncertainties.
The uncertainty of the time lag was estimated using the Flux Randomization and Random Subset Selection (FR/RSS) Monte Carlo method \citep{Peterson_1998}. In each realization, the observed light curves were resampled to account for sampling effects, with approximately 63.3\% of the original data points randomly selected (RSS), and the flux values were randomized by adding Gaussian noise with standard deviations equal to their measured uncertainties (FR). We generated 10,000 such Monte Carlo realizations for both energy bands. For each pair of simulated light curves, the DCF was recalculated and the centroid lag ($\tau_{\rm cen}$) was determined. The centroid was defined as the weighted average of the time lags for which the DCF coefficients exceeded 80\% of the peak value. The resulting centroid-lag distribution was then used to estimate the final lag and its uncertainty: the median was adopted as the best-fit time lag, while the 16th and 84th percentiles were taken as the 1$\sigma$ confidence interval.

To rigorously estimate the significance of the derived correlation peaks and distinguish physical signals from stochastic fluctuations, we performed a Monte Carlo-based analysis described in \citep{2009MNRAS.397.2004A}. 
Considering the data quality and statistical stability of observations, we utilized the observed LAT light curve as a fixed template for this study. Specifically, 1 million synthetic light curves were generated using the algorithm proposed by \citep{2013MNRAS.433..907E}. 
This algorithm ensures that the simulated light curves replicate both the Power Spectral Density (PSD), including its shape and slope, and the Probability Density Function (PDF) of the flux. 
By applying the identical DCF calculation to the ensemble of simulated light curves, we constructed the probability distribution of the DCF values at each discrete time lag. From these distributions, the corresponding quantiles were determined to define the confidence intervals (e.g., 3 $\sigma$ and 4 $\sigma$).

\section{Results} \label{sec:results}

\subsection{MWL light curves}

In this work, we report on a TeV flaring activity of 1ES 1959+650 observed by LHAASO-WCDA in early 2024, and present a multi-wavelength analysis covering the same period. 
Figure ~\ref{fig:MLC}(A) shows the long-term weekly light curve, while Figures ~\ref{fig:MLC}(B)-(F) present multi-wavelength light curves from MJD 60150 to 60520. It can be clearly seen that before 2024, the source was relatively stable; starting from MJD 60324 (15 January 2024), the source entered an active phase. Using a Bayesian method, we partitioned the light curve into different activity states based on their flux levels (see Appendix~\ref{apped:A} Table~\ref{tab:states} for further details).
Three distinct flaring events were identified in the WCDA light curve: flare $f_1$ (MJD 60324.7–60360.6), flare $f_2$ (MJD 60386.6–60399.5), and flare $f_3$ (MJD 60399.5–60441.4). The first triggered event, $f_{1p}$, is part of flare $f_1$. The Fermi-LAT light curve confirms the presence of these TeV flares and shows a high degree of temporal correlation.

%Three distinct flaring events were identified in the WCDA light curve (MJD 60324.7-60360.6, MJD 60386.6-60399.5, MJD 60399.5-60441.4). The first triggered event, which is part of Flare 1. 
%Based on these detections, we defined different flux states (see Figure~\ref{fig:MLC}). 
%The Fermi-LAT light curve shows flaring activity temporally coincident with the TeV events. 
%Following the method described in Appendix \ref{appendix:B}, we find that the temporal coincidence between the flares observed in the two energy bands corresponds to a significance of $4.29\sigma$.

The X-ray light curves also exhibit signs of flaring activity. During Flare 1, both the soft X-rays (2–20 keV, from MAXI) and the hard X-rays (15–50 keV, from Swift-BAT) show signs of flux enhancement, with significant excess during the 1st trigger flare. However, neither MAXI nor BAT had coverage during flare $f_2$ (hereafter ``Hard-lag Flare"). Swift-XRT observations reveal significant flux variability, though they did not cover the TeV-triggered periods. ASAS-SN optical observations in the g-band covered the period of the Hard-lag Flare, but no clear flaring structure was observed. Instead, it appears to lie on a broader, long-term declining plateau.

\subsection{Time lags}
Figure~\ref{fig:DCF} presents the DCF distributions between the GeV and VHE $\gamma$-ray light curves across different temporal intervals. The shaded regions denote the 3 $\sigma$ and 4 $\sigma$ confidence intervals derived from the Monte Carlo simulations described in Section~\ref{simu}.

Focusing on the period of Flare~2, we obtained a characteristic time lag of $\Delta t = 5.0_{-2.1}^{+2.1}$~days after excluding Flare~1 and Flare~3 from the light curves spanning MJD~60300--60500, indicating that the VHE flux variations lag behind the GeV emission. 
Monte Carlo simulations confirm that the significance of this DCF peak reaches $3.4\,\sigma$.
This time delay is consistent with the visual inspection of Figure~\ref{fig:MLC}, where the VHE Hard-lag Flare peaks are delayed relative to the GeV counterpart. 
The chance-coincidence probability was estimated using Monte Carlo randomization tests, in which the VHE flare occurrence times were randomly redistributed within the observational window while preserving the observed flare durations. The resulting probability of temporal coincidence for the GeV and VHE flare activities during Flare~2 is $p = 0.009$, supporting a physical association between the two bands.

%The panel(B) shows the DCF calculated using the full dataset (MJD 60150--60500). The resulting time lag is $\Delta t = 3.0_{-3.0}^{+2.7}$~days, with a statistical significance exceeding $4\sigma$. The decrease in time lag observed in the full-period analysis indicates that the overall correlation represents an ensemble average. This reflects a combination of synchronous variations during the quiescent phase and Flare 1, and the delayed VHE emission observed after Flare 1.
In contrast, the analysis of the full dataset (MJD 60150--60520, Figure~\ref{fig:DCF}(B)) reveals a more complex correlation profile. Rather than a single peak, the distribution exhibits a broad structure that appears to be a superposition of two distinct components: a synchronous signal ($\Delta t \approx 0$ days) and a lagged signal consistent with the delay found in the partial analysis ($\Delta t \approx 5.0$ days). Notably, both components show statistical significance exceeding the 4 $\sigma$ level. This dual-component structure suggests that the full-period correlation is an ensemble average, reflecting both the synchronous variations dominating the quiescent/Flare 1 phases and the delayed VHE emission characteristic of the subsequent activity.

The hard delay phenomenon observed in the Hard-lag Flare is exceptionally rare. Previous studies have reported similar delays only in TeV observations of Mrk 501 \citep{2007ApJ...669..862A}, although those delays occurred on timescales of $\sim$minutes. Therefore, the delay discovered in \tSrc~represents the only known case occurring on a day-scale.

\subsection{Spertral Energy Distributions}

\begin{figure}
\plotone{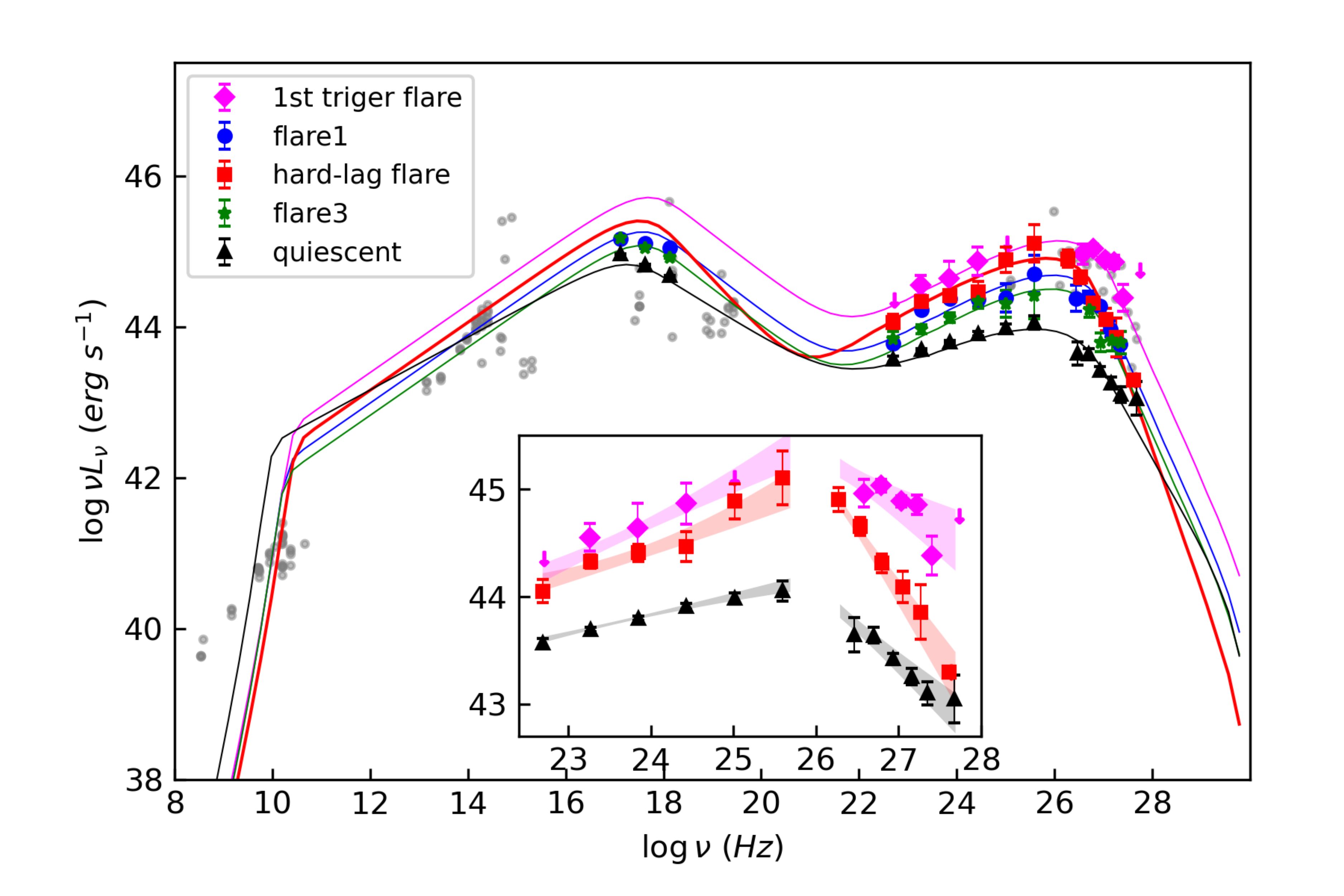}
\caption{Broadband SEDs of 1ES 1959+650. Colored points denote measurements selected for the corresponding activity intervals, whereas gray points show archival measurements; because the multiwavelength coverage is incomplete, the colored data are not all strictly simultaneous. The solid curves are stationary one-zone SSC representations. Flare $f_{2}$ and the quiescent state use separate parameter sets, while the other active states share the adopted intrinsic parameters and differ only in the Doppler factor. These curves are illustrative spectral representations and are not time-dependent fits to the observed GeV-TeV lag.}
\label{fig:SED}
\end{figure}

Based on these defined periods, we further constructed the multi-wavelength SEDs for the different activity states (see Figure~\ref{fig:SED}). 
The SED peaks consistently reside between the GeV and VHE $\gamma$-ray bands. 
For the Swift-XRT and WCDA datasets, the energy spectra were modeled using a standard power-law functional form; notably, the WCDA analysis incorporates the EBL absorption effect to account for extragalactic attenuation. 
During the 1st trigger flare period, the median energy of the highest detected $N_{\text{hit}}$ bin in the WCDA dataset reaches 10~TeV, whereas it extends to 18~TeV during the quiescent state. This difference is due to the significantly longer integration time during the quiescent phase, which allows sufficient photon statistics to accumulate in the highest-energy bins despite the lower instantaneous flux.
In contrast, the Fermi-LAT spectra were characterized by a log-parabola model to accommodate the intrinsic spectral curvature observed in the GeV range.

In the VHE band, the 1st trigger Flare, denoted as Flare $f_{1p}$ in Table \ref{tab:spectral_results}, exhibits the hardest spectrum ($\Gamma =2.48\pm 0.21$), followed by the quiescent state ($\Gamma =2.68\pm 0.13$), while the Hard-lag Flare, denoted as Flare $f_{2}$ in Table \ref{tab:spectral_results}, represents the softest state ($\Gamma =3.16\pm 0.18$). 
In the GeV band (log-parabola fits), the 1st trigger Flare remains the hardest ($\alpha =1.64\pm 0.07$); however, the Hard-lag Flare ($\alpha =1.75\pm 0.06$) appears harder than the quiescent state ($\alpha =1.83\pm 0.02$). This discrepancy in spectral hardness between the GeV and VHE regimes during the Hard-lag Flare suggests a complex shifting of the peak frequency during high-activity periods.

\begin{table}[htbp]
\begin{center}
\caption{Spectral Fitting Results for Different Flux States of 1ES 1959+650.}
\label{tab:spectral_results}
%\begin{tabular}{@{}l|ccc|cc@{}}
\begin{tabular}{c|ccc|cc}
\toprule
& \multicolumn{3}{c|}{\textbf{WCDA}} & \multicolumn{2}{c}{\textbf{LAT}} \\
\textbf{Period} & \textbf{Norm. $F_0$} & \textbf{Spectral Index} & \textbf{TS} & \textbf{Norm. $N_0$} & \textbf{Index} \\
 & ($10^{-12}$ TeV$^{-1}$ cm$^{-2}$ s$^{-1}$) & ($\Gamma$) & & ($10^{-12}$ MeV$^{-1}$ cm$^{-2}$ s$^{-1}$) & ($\alpha$)\\ %\midrule
\hline
Flare $f_1$ & $55.22 \pm6.02$ &  $2.71\pm0.12 $& 245.79 & $7.17 \pm 0.18$ & $1.73 \pm 0.02$ \\

Flare $f_{1p} $ & $166.39 \pm 33.72$ & $2.48\pm0.21$ & 152.16 & $12.90 \pm 1.53$ & $1.64 \pm 0.07$ \\

Flare $f_2$ & $67.16 \pm 7.81$ & $3.16 \pm 0.18$ & 93.25 & $7.93 \pm 0.99$ & $1.75 \pm 0.06$ \\

Flare $f_3$ & $23.15 \pm 4.68$ & $2.75 \pm 0.21$ & 55.13 & $4.58 \pm 0.11$ & $1.73 \pm 0.02$ \\ %\bottomrule
Quiescent & $7.49 \pm 1.07$ & $ 2.68 \pm 0.13$ & 170.57 & $2.10 \pm 0.06$ & $1.83 \pm 0.02$ \\ %\midrule
\hline
\end{tabular}
\end{center}
\begin{flushleft}
\footnotesize 
\textbf{Notes:} $F_0$ is the differential flux at the pivot energy $E_{\text{piv}} = 1$ TeV. Errors are at the 1 $\sigma$ statistical level. The TS represents the Test Statistic of the source detection in each period. The Fermi-LAT spectrum is described by a log-parabola: $dN/dE = N_{0}(E/E_b)^{-(\alpha + \beta {\rm ln}(E/E_b))}$, where $E_b = 1827.53$ MeV is taken from the catalog \citep{2020ApJS..247...33A} and kept fixed as recommended.
\end{flushleft}
\end{table}

\section{Discussion} \label{sec:disc}

Before discussing possible mechanisms, we stress that the following estimates are intended as order-of magnitude consistency checks rather than a unique model identification. The measured lag, the best-fit soft TeV spectrum, and the state-integrated SED do not by themselves establish a specific particle acceleration mechanism. After a long-term quiescent state, 1ES 1959+650 enters an active period starting from roughly 15 January 2024, with three recognized flares, and very high flux variability on a day timescale during the first trigger flare on 6 February 2024. 
% (labeled as the ``Dragon'' flare). 
Broad-band emission is further analyzed from the optical, X-ray, and \textit{Fermi}/LAT GeV bands. Broad-band SEDs are constructed for these flares and the quiescent state, as presented in Figure~\ref{fig:SED}. Unfortunately, not all states have simultaneous broadband data.

\subsection{The hard lag}

\begin{figure}
\plotone{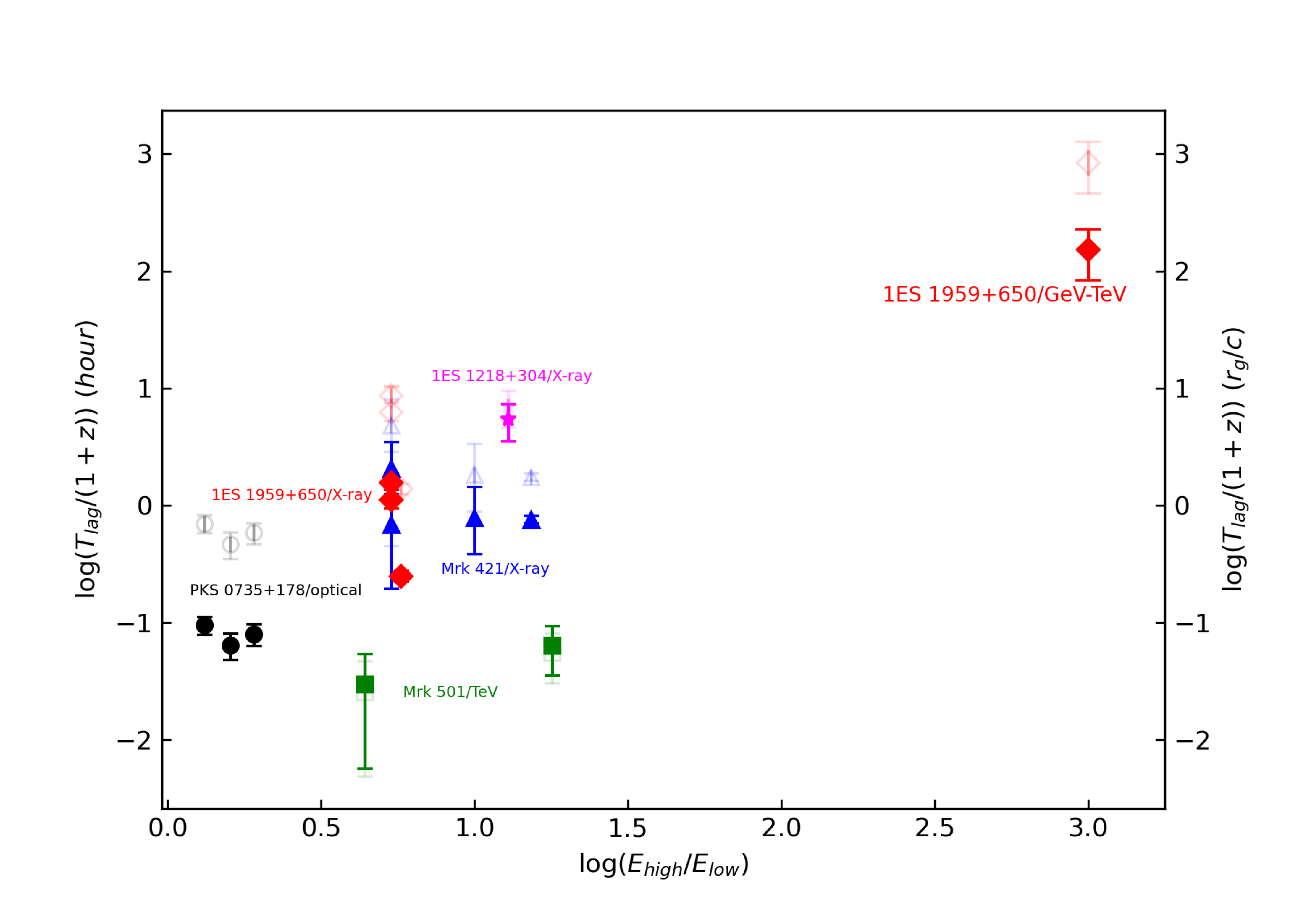}
\caption{Hard lags between band flux variability $(E_{h1},E_{h2})$ and $(E_{l1},E_{l2})$ in blazars. The effective observing frequencies/energies are defined as $E_{high}\equiv\sqrt{E_{h1}E_{h2}}$ and $E_{low}\equiv\sqrt{E_{l1}E_{l2}}$. Data are collected from \citep{2007ApJ...669..862A} for Mrk 501 at TeV band, \citep{2000ApJ...541..153F, 2002MNRAS.337..609Z, 2023MNRAS.524.3797D} for Mrk 421 at X-ray band, \citep{2008ApJ...680L...9S} for 1ES 1218+304 at X-ray band, \citep{2024MNRAS.528.4702M} for PKS 0735+178 at optical band, \citep{2023ApJ...951...94W, 2023MNRAS.524.3797D} for 1ES 1959+650 at X-ray band and this work for 1ES 1959+650 at GeV-TeV band. Considering normalized by different black hole mass, the light open symbols refer lag times in units of $r_{g}/c$ with $r_{g}\equiv GM/c^{2}$ being the gravitational radius and $z$ being the redshift. See context for details.}
\label{fig:hardlag}
\end{figure}

%A particularly interesting feature is revealed in the well-sampled flare~2: we find a lag of the TeV emission behind the GeV emission. The simulations suggest that this delay is statistically significant (at a level of 3.4 $\sigma$, see Figure \ref{fig:DCF}). The phenomenon of a hard lag variability at high frequency lagging behind that at low frequency is very rare. 
A particularly interesting feature is revealed in the well-sampled flare~2: The cross-correlation function peaks at $\Delta t = 5.0_{-2.1}^{+2.1}$ days, with the TeV emission lagging the GeV. 
The phenomenon of a hard lag variability at high frequency lagging behind that at low frequency is very rare. 
%A hard lag of the high-frequency variability behind the low-frequency one, if confirmed, would be a very rare phenomenon.
MAGIC observations of Mrk 501 in 2005 found a flare whose time lag continuously increases with increasing energy difference, reaching $\sim4$ minutes of $1.2$--$10$~TeV lag behind $0.15$--$0.25$~TeV \citep{2007ApJ...669..862A}. To our knowledge, this is the only hard-lag case among robustly reported VHE inter-band. In the X-ray band, 1ES 1218+304 \citep{2008ApJ...680L...9S}, Mrk 421 \citep{2002MNRAS.337..609Z, 2023MNRAS.524.3797D, 2000ApJ...541..153F}, and our target 1ES 1959+650 \citep{2023MNRAS.524.3797D, 2023ApJ...951...94W} were found to exhibit the hard-lag phenomenon, typically on the order of $\sim$hours, and also show increasing lag time with increasing energy difference. Among these sources, Mrk 421 and 1ES 1959+650 are observed to present this phenomenon many times. In the optical band, \citep{2024MNRAS.528.4702M} found PKS 0735+178 to present a hard lag among the $g$, $r$, $i$, and $z_{s}$ bands and also to show increasing lag time with increasing energy difference, with typical lags within $\lesssim10$ minutes. For our TeV-delayed flare, we also tried to find whether there is any similar increase of lag time with increasing energy difference in TeV--GeV bands. We fail to detect it due to the low data significance. For all these robust hard-lag events, we plot them together (illustratively) in Figure \ref{fig:hardlag}. To illustrate the frequency/energy difference between lagged frequencies, a parameter $\log{(E_{\rm high}/E_{\rm low})}$ is defined and shown as the $x$-axis of Figure~\ref{fig:hardlag}. It is clear that the hard-lag time of 1ES~1959+650 at TeV--GeV bands, $\tau_{\rm lag}=5.0_{-2.1}^{+2.1}$ days, is significantly larger than all other known lag times and also has the largest energy difference (TeV relative to GeV). This combination (days-scale lag together with a much wider energy separation) seems to imply the event is particularly intriguing. It may indicate that the dominant origin of the TeV--GeV hard lag is not necessarily identical to that of the previously reported minute--hour cases.

\begin{figure}
\plotone{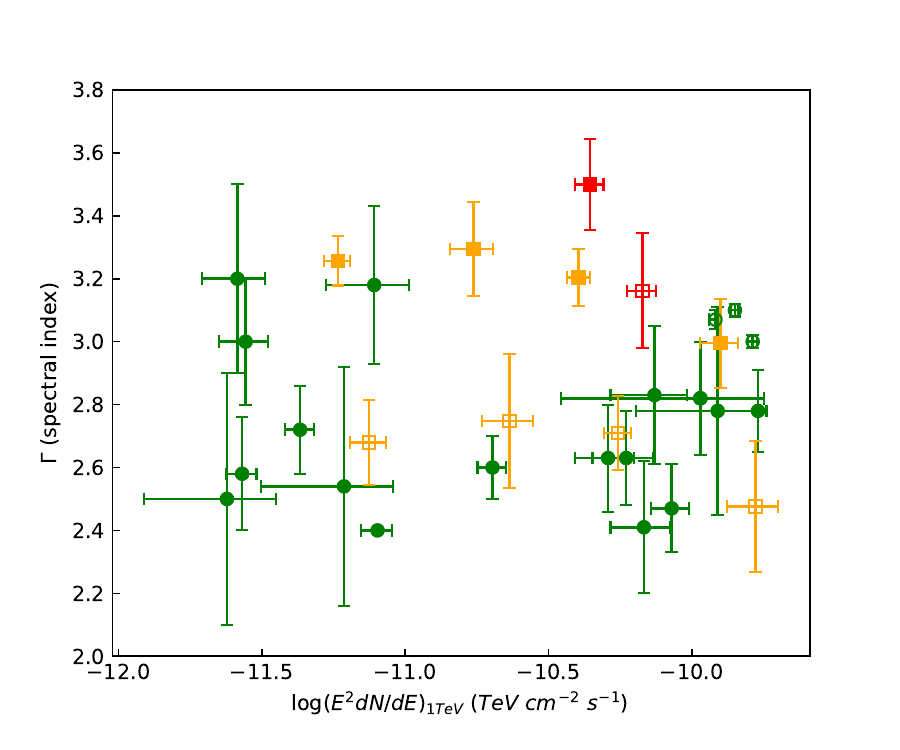}
\caption{TeV spectra versus flux of 1ES 1959+650. The green filled circles are for historical observational spectra collected from \citep{2003A&A...406L...9A, 2005ApJ...621..181D, 2006ApJ...639..761A, 2008ICRC....3.1021H, 2013ApJ...775....3A, 2014ApJ...797...89A} with the green open circles are for EBL corrected intrinsic spectra from MAGIC observations \citep[see][]{2020A&A...638A..14M}.
%with the orange circle being for the ``orphan" flare observed by Whipple at 4 June 2002 \citep{2005ApJ...621..181D}.
The filled orange squares are for LHAASO observations of flares 1, 3, 1st trigger flare %``Dragon" 
and quiescent states, while the filled red square refers to the observational hard-lag flare from this work. The open squares are for the corresponding EBL-corrected intrinsic spectra. See context for details.}
\label{fig:flux-Gamma}
\end{figure}

1ES 1959+650 is very bright at VHE bands and has been observed by all major TeV IACT telescopes, with variable fluxes and spectra. To explore the properties of this hard-lag flare among previous observations, we collect the flux and spectral index of historical TeV observations, combined with LHAASO observed and EBL-corrected intrinsic spectra, all shown in Figure \ref{fig:flux-Gamma} (see the caption for details). It can be seen that the hard-lag flare (the red square in Figure \ref{fig:flux-Gamma}) lies toward the soft end of the historical TeV measurements, with the observed spectral index reaching $\Gamma=3.50\pm0.14$.

Soft-lag variability is more often observed, which is mainly due to the dominance of the cooling process. The opposite ordering, where high-frequency photons arrive later (hard lag), suggests that acceleration-related timescales (or more generally, energization and transport processes) may become important in shaping the observed variability \citep[see e.g.][]{1998A&A...333..452K}. The previously reported minute-hour hard lags have often been interpreted in terms of efficient Fermi-I acceleration \citep[see][]{2007ApJ...669..862A, 2000ApJ...541..153F, 2002MNRAS.337..609Z, 2023MNRAS.524.3797D, 2008ApJ...680L...9S, 2024MNRAS.528.4702M, 2023ApJ...951...94W, 2023MNRAS.524.3797D}, consistent with theoretical expectations \citep{1998A&A...333..452K, 2011A&A...525A..40M} in the regime where electrons approach their maximum energies such that $t_{\rm acc}\lesssim t_{\rm cooling}$. The Fermi-I acceleration timescale of a relativistic shock, $T_{\rm acc,I}\approx \eta\gamma m_{e}c/eB$, is typically much faster than the observed day-scale lag, unless the scattering mean free path is very large. In contrast, if the variability involves a slower effective energization channel (e.g., stochastic acceleration in turbulence under certain conditions), the characteristic timescale can be substantially longer and may become comparable to the observed day-scale TeV--GeV lag.

In this context, a stochastic acceleration (Fermi-II) interpretation is one plausible route to connect the observed lag amplitude to reasonable jet conditions, without requiring extremely rapid energization. In the hard-sphere limit, the energy-diffusion coefficient is $D_{\gamma\gamma}\propto\gamma^{2}$, for which the characteristic stochastic acceleration time is approximately energy-independent, $T_{\rm acc,II}\approx\gamma^{2}/(2D_{\gamma\gamma})$. This implies that the time to accelerate electrons from $\gamma_{1}$ to $\gamma_{2}$ is logarithmic, $\Delta t_{1\rightarrow2}= t_{\rm acc}\ln\left(\gamma_{2}/ \gamma_{1}\right)$, which could be compatible with our TeV hard-lag flare given that the TeV and GeV variability timescales are of the same order as the lag time. Using the representative SSC parameters (see below), the magnetic field $B\sim0.02$~G and the Doppler beaming factor $\delta\sim15$, we adopt $\gamma_{G}\sim3\times10^{4}$ and $\gamma_{T}\sim3\times10^{5}$ as characteristic, rather than unique, Lorentz factors for the broad GeV and TeV bands. For mildly magnetized turbulence \citep[Alfv\'{e}n velocity $v_{A}\approx0.1c$,][]{2003A&A...397...15T, 2008ApJ...681.1725S, 2011JCAP...12..010M}, the Fermi-II acceleration time reads $t_{\rm acc,II}\approx \left(c/v_{A}\right)^{2}\lambda/c\sim 50$ days (in comoving frame), which gives the observed lag time between GeV and TeV emissions $\tau_{\rm lag}\sim7$--$8$ days, similar to what we observe for the TeV hard lag in 1ES~1959+650. For these parameters, the dominated cooling time of $\gamma_{T}$ is $t_{\rm cooling}\approx 3 m_{e}c/(4\gamma_{T}\sigma_{T}U_{B})\sim75$ days, which is just slightly larger than the acceleration timescale. If acceleration and total losses become comparable near $\gamma_{T}$, the electron spectrum may steepen, qualitatively consistent with the soft TeV spectrum.

At the same time, the interpretation is not unique. A colliding-shell internal shock could in principle form a forward shock and a reverse shock that dominate different $\gamma$-ray bands and thereby produce a hard lag. However, the present data do not constrain the relative magnetic fields, particle distributions, or radiative efficiencies of the two shocked regions well enough to test this scenario quantitatively. More generally (and more briefly), other possibilities that could contribute to a TeV-lag signature include multi-zone/stratified emission (with different characteristic acceleration or radiative efficiencies across zones), time-dependent SSC effects where the seed photon field and the highest-energy electrons evolve on comparable timescales, or evolving opacity/escape that delays the emergence of the highest-energy component. These scenarios are not developed here, but they motivate treating the TeV--GeV hard lag as a diagnostic that is both rare and potentially discriminating, rather than as evidence for any single mechanism by itself. As discussed above, for this hard-lag flare, we also attempted an energy-resolved lag analysis by subdividing the LAT and WCDA data. However, the reduced photon statistics in the subbands meant that no meaningful lag-energy relation could be measured. The GeV and TeV peaks may arise from two distinct dissipation events during a common episode of central-engine activity. For example, two shell pairs ejected at different times could collide at different radii, where different physical conditions produce GeV- and TeV-dominated emission, respectively. In this picture, the measured lag would characterize the separation between the two dissipation events rather than a particle-acceleration timescale.

\subsection{One-zone SSC representations}

Figure \ref{fig:SED} shows the broadband SEDs of 1ES 1959+650, which are modeled by a one-zone SSC model considering EBL absorption correction \citep{2021MNRAS.507.5144S}. A homogeneous and isotropic emission region is assumed, with a sphere of radius $R$, a uniform magnetic field with strength $B$, a broken power-law electron energy density distribution $N(\gamma)=N_{0}\gamma^{-p_{1}}$ at $\gamma_{\rm min}\leq\gamma\leq\gamma_{0}$ and $N(\gamma)=N_{0}\gamma_{0}^{p_{2}-p_{1}}\gamma^{-p_{2}}$ at $\gamma_{0}\leq\gamma\leq\gamma_{\rm max}$, and the Doppler beaming factor $\delta$; see \cite[][]{2017ApJ...842..129C} for details. These stationary SSC calculations are used only to provide representative model parameters for $f_{2}$ and characteristic electron Lorentz factors for the order-of-magnitude estimates in Section 4.1; they are not intended to reproduce the temporal lag. Because the GeV and TeV peaks of $f_{2}$ occur at different times and no simultaneous X-ray measurements are available, its state-integrated SED should not be interpreted as an instantaneous or uniquely constrained electron distribution. Considering the different observational characteristics of the hard-lag flare and the quiescent state, we use separate representative parameter sets for $f_{2}$ and the quiescent state, as listed in Table 2. The other active-state SEDs can be represented phenomenologically with a common set of intrinsic parameters while varying only the Doppler factor. Given the degeneracy of one-zone SSC modeling and the incomplete simultaneity of the broadband data, this interpretation should not be regarded as unique. For blazars with superluminal motions at VLBI scales, jet velocities often change among different epochs \citep[e.g.,][]{2001ApJS..134..181J}. These modeled Doppler beaming factors are consistent with those derived from the $\gamma$-ray opacity constraint \citep[see e.g.,][]{1995MNRAS.273..583D, 1999A&A...349...11A}, where  $\tau_{\gamma\gamma}=\frac{(\nu L_{\nu})\sigma_{T}E}{20\pi m_{e}^{2}c^{6}\delta^{6}\Delta t_{obs}}<1$ requires $\delta\gtrsim10$, given insensitive dependence $E\sim 3$ TeV, $\nu L_{\nu}\sim 10^{45}$ erg $s^{-1}$ roughly near the corresponding target-phton energy, and the variability timescale $\Delta t_{\rm obs}\sim1$ day.

If the observed TeV lag reflects the time required to energize particles from the GeV to TeV emitting populations while the emitting plasma propagates relativistically along the jet, the later TeV-dominated emission would arise farther downstream than the earlier GeV-dominated emission, with an associated geometry-dependent propagation distance $\Delta r\sim 1\left(\frac{\Gamma}{15}\right)^{2 }\left(\frac{\Delta\tau_{lag}}{5\ \mathrm{days}}\right)$ pc in the AGN frame. When considering high-resolution VLBI observations of 1ES 1959+650, it seems that the relativistic jet motion discussed above conflicts with the $\sim$pc-scale measurements, where VLBI pattern speeds are subluminal or stationary on parsec scales \citep{2010ApJ...723.1150P}. This may be due to a spine/layer nature of the jet \citep{2005A&A...432..401G, 2017ApJ...842..129C}, in which the VHE emission is produced in a faster spine nested within a slower sheath, where the high-resolution radio emission is produced \citep{2010ApJ...723.1150P}.

\begin{table}
\tabletypesize{\tiny}
\setlength{\tabcolsep}{4.0pt}
\caption{Representative one-zone SSC parameters}
\begin{tabular}{@{}lcccccccccccccc@{}}
\hline
\hline
States  & $\delta$  & $B$ (Gs) & $R$ (cm)  & $\gamma_{0}$  & $p_{1}$ & $p_{2}$  & $N_{0}$\\
\hline
Flare $f_1$ & 15 & 0.015  & $9.6\times10^{16}$ & $6.3\times10^{5}$   & 2.1 & 4.0   & 185         \\
Flare $f_{1p}$ & 20 & 0.015  & $9.6\times10^{16}$ & $6.3\times10^{5}$   & 2.1 & 4.0   & 185         \\
Flare $f_2$ & 15  & 0.023 & $9.6\times10^{16}$ & $5.0\times10^{5}$ & 2.1 & 4.4   & 329      \\
Flare $f_3$ & 14 & 0.015  & $9.6\times10^{16}$ & $6.3\times10^{5}$   & 2.1 & 4.0   & 185         \\
Quiescent & 11.5 & 0.008  & $4.9\times10^{17}$ & $6.7\times10^{5}$   & 2.3 & 3.8   & 153         \\
\hline
\end{tabular}
\\
\tablecomments{For the active-state parameters, the observer-frame light-crossing times are, $t_{lc,obs} = (1+z)R/(c\delta)\sim2-3$ days; the stationary models describe the multi-day flare envelopes rather than the shortest daily substructure.}
\label{tab:SEDparameters}
\end{table}

\section{Summary}
We present a multi-wavelength study of the TeV activity of 1ES~1959+650 in 2024 triggered by the LHAASO-WCDA real-time monitoring/alert system. A Bayesian-block analysis identifies three TeV flares, including the 1st trigger flare on 2024 February 9 and two subsequent flaring episodes. 
%\textcolor{red}{The highest photon energy reaches $\sim18$ TeV. The key result is a statistically significant day-scale hard lag during the second triggered event: the TeV emission lags the GeV emission by $\Delta t \simeq 5.0_{-2.1}^{+2.1}$ days, with a Monte-Carlo significance \textcolor{red}{reaching 3.4 $\sigma$} based on red-noise simulations that reproduce both the PSD and PDF of the observed light curves.} 
%The key result is a \textcolor{red}{significant correlation between GeV and TeV emission during the second triggered event, with a Monte Carlo significance of 3.4~$\sigma$ based on red-noise simulations that reproduce both the PSD and PDF of the observed light curves. The cross-correlation function peaks at a time delay of $\Delta t \simeq 5.0_{-2.1}^{+2.1}$ days.}
The DCF analysis suggests a peak time delay of $\Delta t = 5.0_{-2.1}^{+2.1}$ days. Red-noise simulations of uncorrelated light curves indicate a correlation significance of $3.4\,\sigma$ at this lag.
The hard-lag flare shows the softest TeV spectrum among the WCDA flares (the observed spectral index $\Gamma_{\rm obs}\simeq3.5\pm0.14$ and EBL corrected intrinsic one $\Gamma_{\rm int}\simeq 3.16\pm0.18$), and its SED can be reproduced with a one-zone SSC model with parameters distinct from the other flux states, while the remaining flares/states can be described with broadly similar jet parameters but different beaming factors. The unusual combination of a TeV--GeV hard lag on a day-timescale and the accompanying soft TeV spectrum provides a rare and potentially discriminating diagnostic of the underlying variability physics. The observed lag and spectral behavior are compatible with acceleration-related evolution of the emitting electrons, but the interpretation is not unique; multi-zone/stratified emission, single-collision FS/RS configurations, and distinct dissipation events within a common activity episode remain viable. Future strictly simultaneous X-ray--GeV--TeV monitoring with dense cadence will be crucial for measuring energy-dependent lags, discriminating among these scenarios, and better constraining the geometry and dynamics of the $\gamma$-ray-emitting region.

\section{Acknowledgments}
The LHAASO Observatory, including its detector systems, was designed and constructed by the LHAASO project team and is operated and maintained by the LHAASO operations team. We sincerely thank all members of both teams, with special appreciation for those who work year-round at the LHAASO site at an altitude exceeding 4,400 meters. Their sustained dedication ensures the reliable operation of the detector systems and essential infrastructure, including the power supply.

We sincerely acknowledge the Chengdu Management Committee of Tianfu New Area for its sustained financial support of research based on LHAASO data. We also thank the National High Energy Physics Data Center for providing the computing resources and data services that made the analysis in this work possible.

\subsection{Funding}
This work is supported in China by National Key R\&D program of China under the grant 2024YFA1611403, 2024YFA1611401, 2024YFA1611402, 2024YFA1611404;
%This work was supported by the National Key R\&D Program of China (Grants No. 2024YFA1611401--2024YFA1611404); 
the National Natural Science Foundation of China (Grants No. 12393851--12393854, 12173039, 12205314, 12105301, 12305120, 12261160362, 12105294, U1931201, and 12375107); the Department of Science and Technology of Sichuan Province (Grant No. 24NSFSC2319); the Project for Young Scientists in Basic Research of the Chinese Academy of Sciences (Grant No. YSBR-061); Youth Innovation Promotion Association CAS (No. 2023275); and, in Thailand, by the National Science and Technology Development Agency (NSTDA) and the National Research Council of Thailand (NRCT) under the High-Potential Research Team Grant Program (Grant No. N42A650868).

\subsection{Author Contributions}
All authors contributed equally to this work. Specifically, G.M.~Xiang and M.~Zha performed the LHAASO data analysis, while J.N.~Zhou analyzed the multi-wavelength data. L.~Chen provided theoretical modeling and interpretation. All other authors participated in various aspects of the data analysis, including data acquisition,  data processing, data quality assessment, detector calibration, event reconstruction, and simulations, and provided comments on the manuscript.

\appendix

\section{LHAASO-WCDA Analysis}
\label{apped:A}
The LHAASO instrument features a wide field of view (FoV) spanning a solid angle of approximately $2.2$~sr, exhibiting peak sensitivity to sources within the declination range of $-20^\circ$ to $+80^\circ$. Due to the Earth's diurnal rotation, any celestial position within this declination band transits through the FoV once per sidereal day, enabling continuous daily monitoring of candidate sources.

However, as a ground-based air shower array, the effective observation window for a specific source is concentrated within several hours around its culmination (transit). To mitigate potential observational biases arising from seasonal variations or solar time effects, we binned the background-estimated sky maps based on the specific transit time of 1ES 1959+650 within the LHAASO FoV. This procedure ensures that each sampling point is analyzed under a unified temporal reference frame, maintaining consistent detection efficiency and exposure across all intervals. Consequently, this approach yields a measurement of the unbiased daily flux, providing a robust basis for long-term variability studies. The fundamental analysis unit was defined as 1 transit, and sky maps for longer durations were generated by integrating these 1-transit intervals sequentially.

The analysis of each temporal bin was performed using a 3D maximum likelihood fitting method. We defined a circular Region of Interest (ROI) with a \(3^{\circ }\) radius centered at (\(\text{R.A.}=300^{\circ },\text{Decl.}=65.15^{\circ }\)). Given the extragalactic nature of 1ES 1959+650, the source was modeled as a point source.
The intrinsic energy spectrum of the source is assumed to follow a power-law function. However, very-high-energy (VHE) gamma rays from extragalactic sources are significantly attenuated by the Extragalactic Background Light (EBL) via pair production ($\gamma \gamma \to e^+ e^-$). Consequently, the observed photon flux arriving at Earth is modeled as:
\begin{equation}
    \frac{dN}{dE} = F_0 \left( \frac{E}{E_{\text{piv}}} \right)^{-\Gamma} \times \exp[-\tau(E, z)]
\end{equation}
where:
\begin{itemize}
    \item $F_0$ is the spectral normalization parameter (flux at the pivot energy);
    \item $E_{\text{piv}}$ is the pivot energy, strategically chosen to minimize the correlation between $F_0$ and the spectral index;
    \item $\Gamma$ is the intrinsic spectral index;
    \item $\tau(E, z)$ denotes the optical depth for a gamma-ray with observed energy $E$ from a source at redshift $z$. 
\end{itemize}
In this work, we adopt the EBL opacity model from \citep{2021MNRAS.507.5144S}, which provides an updated determination of the evolving EBL spectral energy distribution based on multiwavelength galaxy surveys.
For 1ES 1959+650, the redshift is set to z=0.048 \citep{1996ApJS..104..251P}.

We first determined the source position and spectral index $\Gamma$ from the full long-term dataset, where high photon counts ensure robust constraints. These parameters were then fixed in the transit-by-transit fits, leaving only the flux normalization as a free parameter. By reducing the degrees of freedom, this procedure yields stable flux measurements even for transits with limited exposure time.

To objectively identify the change points between different flux states, we applied the Bayesian Blocks algorithm \citep{2013ApJ...764..167S} to the transit-binned light curve of 1ES~1959+650. This non-parametric method partitions the time-series data into blocks of constant flux, effectively capturing significant intensity variations while filtering out statistical noise. In our analysis, the input for each sampling point consisted of the flux measurement $F_i$ and its uncertainty $\sigma_i$. Within each identified block, the representative constant flux was calculated using a weighted average with weights $1/\sigma_i^2$. To control the sensitivity of the algorithm, we set the false positive probability to $P_0 = 0.05$.

The algorithm identified 8 distinct flux states, whose time intervals and flux characteristics are summarized in Table~\ref{tab:states}. In this study, State~1 is defined as the quiescent state. A flaring state is identified when the flux exceeds twice the quiescent level ($F_{1-20} > 2.54 \times 10^{-12}$ cm$^{-2}$ s$^{-1}$). Based on this criterion, five flaring states are identified: States~2, 3, 4, 6, and 7. The consecutive States~2--4 collectively constitute the first major flaring episode (Flare~1), within which State~3---exhibiting the highest flux among all identified states---is designated as Flare~1p (the peak component). States~6 and 7 are defined as Flare~2 and Flare~3, respectively. These three distinct flaring periods, along with the peak substructure of Flare~1, are clearly marked in the multi-wavelength light curves presented in Figure~\ref{fig:MLC}, providing the basis for our subsequent state-resolved spectral analysis.

\begin{table}[htbp]
\centering
\caption{Temporal Segmentation of 1ES 1959+650 Flux States Identified by the Bayesian Blocks Algorithm.}
\label{tab:states}
\begin{tabular}{@{}clcccc@{}}
\toprule
\textbf{Block ID} &   \textbf{State} &  \textbf{Start Time} & \textbf{End Time} & \textbf{Flux $F_{1-20}$} & \textbf{Flux Error $\Delta F_{1-20}$} \\
 & & (MJD) & (MJD) & ($10^{-12}$ cm$^{-2}$ s$^{-1}$) & ($10^{-12}$ cm$^{-2}$ s$^{-1}$) \\ \midrule
\hline
1 & low /(quiescent) & 59281.592 & 60324.736 & 1.27  &  0.21\\
 2 & high /($f_{\rm 1,rise}$)  & 60324.736 & 60346.677 & 13.15  &  1.56\\
 3 & high /($f_{1p}$) & 60346.677 & 60348.671 & 62.72 &  6.58\\
 4 & high /($f_{\rm 1,decay}$) & 60348.671 & 60360.639 & 11.07 &  2.02\\
 5 & low & 60360.639 & 60386.568 & 1.20  &  1.30\\
 6 & high /($f_2$) & 60386.568 & 60399.532 & 16.08 &  2.12\\
 7 & high /($f_3$) & 60399.532 & 60441.417 & 5.18  &  1.02\\
 8 & low & 60441.417 & 60521.199 & -0.03 &  0.68\\ \bottomrule

\hline
\end{tabular}

\textbf{Notes:}{The reported flux values $F_{1-20}$ are the integrated photon flux in the 1--20 TeV energy range. The time intervals are defined by the change points detected with a 5\% false positive threshold.}
\end{table}

\bibliography{1es1959}
\bibliographystyle{aasjournal}



\end{document}